\documentclass[useAMS,usenatbib,onecolumn]{mnras}

\usepackage{latexsym}
\usepackage{epsfig}
\usepackage{amssymb}
\usepackage{amsmath}
\usepackage{color}
\usepackage{mathtools}

\newcommand{\ba}{\begin{eqnarray}}
\newcommand{\ea}{\end{eqnarray}}
\newcommand{\be}{\begin{equation}}
\newcommand{\ee}{\end{equation}}

\newcommand{\vn}{{\mathbf{n}}}
\newcommand{\bn}{\mathbf{n}}

\begin{document}

\title{A signature of EeV protons of Galactic origin}

\author[P.\ G.\ Tinyakov, F.\ R.\ Urban, D.\ Ivanov, G.\ B.\ Thomson, A.\ H.\ Tirone]{P.\ G.\ Tinyakov$^{1}$\thanks{petr.tiniakov@ulb.ac.be}, F. R. Urban$^2$\thanks{federico.urban@kbfi.ee}, D.\ Ivanov$^3$\thanks{ivanov@cosmic.utah.edu}, G.\ B.\ Thomson$^3$\thanks{thomson@physics.utah.edu}, A.\ H.\ Tirone$^4$\thanks{alicia.tirone@cosmic.utah.edu}\\
$^{1}$Service de Physique Th\'eorique, Universit\'e Libre de Bruxelles, CP225, Boulevard du Triomphe, B-1050 Brussels, Belgium\\
$^{2}$National Institute of Chemical Physics and Biophysics (KBFI/NICPB), R\"avala 10, 10143 Tallinn, Estonia\\
$^{3}$Department of Physics and Astronomy, University of Utah, 115 S.\ 1400 East, Salt Lake City, UT 84112, USA\\
$^{4}$White Station High School, 514 S.\ Perkins Road, Memphis, TN 38117, USA\\
}

\date{\today}

\maketitle

\label{firstpage}

\begin{abstract}
We investigate signatures that would be produced in the spectrum and sky
distribution of UHECR by a population of the Galactic sources of
high-energy protons in the energy range around 1~EeV, i.e., around the
diffusive-to-ballistic transition. In this regime, the CR flux has to
be calculated numerically. We employ the approach that consists in
backtracking anti-protons from Earth through the Galaxy and
integrating the source emissivity along the trajectory. This approach
makes evident two generic features of the transition region: sharp
increase of the total flux as the energy decreases across the
transition region, and its strong anisotropy (appearance of a bright
compact spot) all the way until the onset of the diffusive regime. We
then discuss and compare several methods to experimentally detect or
constrain these features. We find that a few percent admixture of the
Galactic protons can in principle be detected by the current UHECR
experiments. 
\end{abstract}



\section{Introduction}\label{sec:intro}

The sources of the high-energy end of the cosmic ray (CR) spectrum have thus
far eluded all attempts for their identification. It is widely believed that
ultra-high energy CRs (UHECRs) with rigidities well beyond 10~EV are of
extra-Galactic origin because the Galactic magnetic field (GMF) cannot confine
them within the Milky Way and because no strong anisotropy associated with the
Galaxy has been observed. For similar reasons, at energies well below 1~PeV
the CRs flux should be dominated by Galactic sources; however, the exact
spectral location of the transition between the two is unknown.

Of particular interest are CRs in the rigidity range from 0.1~EV to 10~EV as
in this region the propagation properties of CRs in the Galactic magnetised
medium change from semi-diffusive to ballistic. This band also harbours two
remarkable yet unexplained spectral features: the Second Knee at about
0.4~EV, where the spectrum softens, thence the ``knee'' appellative; and a
hardening at about 3~EV, which has been therefore christened the Ankle.

There are several competing explanations for the physics behind these two
features~\cite{Nagano:2000ve,Allard:2005cx,Kampert:2014eja}, generally involving either specific features of UHECRs sources, or some physical change in propagation properties. Either feature could be due to an intrinsic limit in the
acceleration capabilities of Galactic sources, and show the transition between
Galactic and extra-Galactic fluxes. Alternatively, the Second Knee could arise
from a change in the diffusion properties of the inter-stellar medium, such as
an drop in confinement time, while the Ankle could be the result of the
interaction between extra-Galactic UHECRs and microwave background photons.

To discriminate between these different possibilities, we study the flux from Galactic sources making use of the fact that, whatever they are, their distribution in space must follow the structure of the Galaxy, i.e., concentrate in the Galactic disk. As we will argue, this property alone results in a characteristic flux shape that, if present, can be readily identified in the data. Thus, although in what follows we work with a concrete source model for which the distribution is known, the results are more general. The presence of generic and easily discernible features in the flux distribution from Galactic sources is the main point of this paper. 

The most interesting effects, as we will see, arise (and are best detected) around 1~EV to 3~EV, where most likely CRs are protons, see~\cite{Abbasi:2009nf,Jui:2011vm,Dedenko:1900fg,Abraham:2010yv}; for this reason, and for simplicity, in such a way that the physics behind our numerical results is most clear and understandable, in what follows we will focus on proton propagation thoughout --- we will thus speak of energy in place of rigidity: since the essential physics in this work is the propagation of UHECRs in a given magnetic field, which depends only on particle rigidity, our approach here is portable to other compositions, once the energy is appropriately rescaled.  In this case, following individual CRs through the GMF is necessary since as we approach the EeV range (from above) the trajectories of some CR primaries begin to show hints of diffusive propagation --- the Larmor radius of a 1~EeV proton in a 1~$\mu$G MF is about 1~kpc. One can rely neither on a purely diffusive approach, nor neglect the deflections altogether.

From our simulations in the transition energy region a general picture emerges. 
\begin{enumerate}
  \item At high ($\gg1$~EeV) energies, the trajectories are mostly ballistic so that the flux roughly follows the source distribution, which is modelled around the shape of the Milky Way.  The highest flux is typically in the direction of the Galactic centre (GC), where the largest number of sources is located.
  \item As we approach the EeV range the trajectories become semi-diffusive, that is, CRs spin around the field lines in the regions with a stronger GMF.  A special direction appears around which the flux is strongly enhanced.  The appearance of this region, typically of the size of several tens of degrees, is linked with the structure of the GMF in the vicinity of the Earth. The flux thus becomes strongly anisotropic.  This effect very generically arises for any order $\mu$G GMF following our Galactic arm: it does not depend on the details of the GMF model, nor it does on the source distribution (provided that sources are localised in the disk).
  \item The flux enhancement is not a re-focussing of existing flux; on the contrary, the total flux \emph{grows} as the energy decreases, on top of the actual injection spectrum. In other words: the modification factor~\cite{Berezinskii} increases at lower energies.
  \item As we proceed further down ($\ll1$~EeV) in energy, the total flux continues to grow (although less steeply); the random GMF becomes more relevant and the region from which the highest flux is collected expands as more and more trajectories become (semi) diffusive, finally covering the whole sky. 
\end{enumerate}

The overall enhancement of the flux in the transition to the diffusive regime is well known --- see~\cite{Aloisio:2004fz,Lemoine:2004uw,Berezinsky:2007kz,Kotera:2007ca,Globus:2007bi,Aloisio:2008tx,Giacinti:2011ww} for some previous works on this topic. The new observation, however, is that this transition happens through a strongly anisotropic stage, resulting in a signature that is easily identifiable in the data, if present. 

The rest of the paper is organised as follows: in Sec.~\ref{sec:met} we define the source distribution and the GMF, and explain our method.  Sec.~\ref{sec:res} contains the results of our procedure; results which we analyse and comment in Sec.~\ref{sec:disc}.  We conclude in Sec.~\ref{sec:conc}.

\section{Method}\label{sec:met}

In order to study the features of a CRs flux which originates from a given local, e.g., Galactic, distribution of sources, we shoot (anti)CRs from the Earth in all directions, and backtrack them through the GMF until they leave the source region and the Milky Way itself.  We then calculate the expected flux in each direction to obtain flux maps and spectra due to Galactic sources.  In this section we detail all these steps.

\subsection{Source distribution}

We want to understand how CRs from a distribution of Galactic sources would look like.  While the details of the source distribution are not really important for the physical picture we arrive at, to be concrete we choose to work with the distribution of pulsars in the Galaxy, which we take from~\cite{Lorimer:2003qc}.  This distribution has decaying radial profile
\be
\rho(r) = r^p \exp(-r/\sigma_r) \, , \quad p=2.35 \, , \, \sigma_r=1.528\;\text{kpc} \, , \label{distror}
\ee
where $r$ is the polar radius from the GC.  In the vertical direction $z$ we instead take the gaussian profile
\be
\rho(z) = \exp(-z^2/2\sigma_z^2) \, , \quad \sigma_z=50\;\text{pc} \, . \label{distroz}
\ee
Finally, we set the source region to zero outside a sphere of 15~kpc. We assume that all sources are identical and emit CRs isotropically. 

Our conclusions do not rely on the details of this particular source distribution; it is enough that the sources are sufficiently numerous (see Sec.~\ref{ssec:flu} below), and concentrated in the Galactic disc. For example, the distribution of supernova remnants as described in~\cite{Green:2013qta}, with radial profile $n(r) = (r/R_\odot)^{0.7} \exp[-3.5(r-R_\odot)/R_\odot]$, as employed in the study~\cite{Giacinti:2014xya}, will give an identical qualitative picture, and a very similar quantitative one.

With such generic assumptions it is also not important whether the sources are transient or not. A quantitative statement will be worked out in the next section.

\subsection{Flux calculation}\label{ssec:flu}

A straightforward way to calculate flux from a given distribution of sources is to propagate CRs emitted isotropically from each source and ``measure'' the flux at the location of the Earth. Despite this method being applicable in all cases, it however becomes very quickly computationally intensive once the number of sources exceeds several hundreds. For this reason, we use this method for cross-check purposes only. 

When the source distribution can be approximated as steady and continuous, another method is more efficient. It is based on the observation that the flux from a given direction can be obtained by launching in that direction an anti-particle and calculating the {\em linear} integral of the source emissivity (density times the luminosity of a single source, assuming sources are identical) along the trajectory. This method has been used previously, e.g., in Refs.~\cite{Giacinti:2011ww,Lee:1995ki,Karakula:1972na}, and it is possible to demonstrate its validity analytically. Apart from the computational advantage, it gives a useful intuitive insight into the flux distribution over the sky.

The backtracking method assumes that the source distribution is continuous and constant at least on the timescales of CRs propagation within the Galaxy; it also assumes that the sources are steady. These conditions, however, can be relaxed: the sources can be discrete and transient, provided their density and the birth rate are large enough. More quantitatively, for this method to give a meaningful result the following condition is needed. As we will see shortly, the region of the flux enhancement corresponds to (back-tracking) trajectories that spend considerably more time in the Galactic disk than they would if they were propagating on a straight line. Together, these trajectories span a certain volume $V$ within the disk. For a continuous approximation to make sense, the number of sources that are active in a given moment of time in the volume $V$ should be much larger than one. Denoting the density of sources $\rho$, the lifetime of a source $t$ and the age of the Universe $T$, this condition reads
\begin{equation}
\frac{t}{T} V \rho \gg 1.
\label{eq:validity_cond}
\end{equation}
The volume $V$ is a sizable fraction of the disk. Taking $V \sim {\rm kpc}^3$, for steady sources we get therefore $\rho \gg 1~{\rm kpc}^{-3}$. For comparison, the density of pulsars in the disk is at least $10^3~{\rm kpc}^{-3}$.

The results presented below are obtained with the backtracking method. However, we have explicitely checked that they are reproduced in the brute force calculation in case of a relatively small number of sources, for which the latter is still possible to carry out.

\subsection{The Galactic magnetic field}

\paragraph*{The regular field}

The second step is the choice of the GMF model.  In all our simulations we employed the large-scale field of~\cite{Pshirkov:2011um}.  This model consists of two components of the GMF, a disk field and a halo field.  The disk field has both radial and azimuthal components and roughly follows the spiral structure of the Galaxy; of the two possible versions of the model, dubbed ``axisymmetric'' and ``bisymmetric'', respectively, we choose the former here as it provides the best fit to the data, and is compatible with the general structure preferred by other recent models (see~\cite{Haverkorn:2014jka} for an overview). The best-fit overall magnitude of the disk field is $B^D_0 = 2\;\mu$G.

The halo field instead is purely toroidal, and the direction of the field reverses across the disk.  The North and South field strength best-fit values are $B_{0,n}^H = 4\;\mu$G and $B_{0,s}^H = - 2\;\mu$G, and the field peaks vertically at 1.3~kpc.  We include all the details in Appendix~A.

In our analysis we have also tested different models, always with very similar 
results. For instance, if we adopt the ``bisymmetric'' model instead, where 
the field is somewhat stronger in the Southern hemisphere, our results are 
even more pronounced. The reason is that within our formalism, the qualitative 
picture that emerges is a trivial consequence of the trapping of charged 
particles in the magnetic field of the disk where most of the sources are 
concentrated. The predictions are hence very robust against the details of the 
model, and are intimately related to the fact that we live in (at the edge of) 
a spiral arm with a regular coherent MF of several $\mu$G and field lines that 
roughly follow the Galactic arm.  The poorly known halo field is, in particular, practically irrelevant (see below Sec.~\ref{ssec:met}).  In the case of the model of~\cite{Jansson:2012pc} we have, consistently, also observed a similar behavior.

\paragraph*{The turbulent field} 

The model of~\cite{Pshirkov:2011um} does not contain any prescription for the 
random field; according to the recent review~\cite{Haverkorn:2014jka}, the 
small-scale field can also be split into a disk and a halo components, with 
different strengths and coherence lengths. The disk component has a reference 
strength $\hat B^D_0$ of a few $\mu$G (we take 5~$\mu$G), and a Kolmogorov 
spectrum with largest coherence length $l_c^D=3$~pc. The halo field instead 
has a much reduced strength of about $\hat B^H_0 = 1\;\mu$G or less, but the 
spectrum (albeit possibly flatter than Kolmogorov) extends out to 
$l_c^H=100$~pc. Both disk and halo fields follow everywhere the structure of 
the regular field. 

The details of the random field (both in the disk and halo) are not very
relevant: its role is merely to blur/jag the picture, so whatever parameters
describe this field, only their particular combination that determines the
strength of the blurring is finally important. It was found
in~\cite{Tinyakov:2004pw,Pshirkov:2013wka,Beck:2014pma} that the overall
average deflections caused by the random field are limited to a fraction of
those propelled by the regular component. In our simulations the strength of
the random field is position-dependent and proportional to the value of the
regular field; we choose its parameters in such a way that, in the ballistic
regime, the random deflections are approximately 1/4th of the regular ones for
the trajectories crossing the Galactic disk at the Earth location. As we will
see below, this random field is not so strong as to disrupt the trajectories
delineated by the dominant smooth GMF.

\section{Simulations and results}\label{sec:res}

\subsection{Simulations}\label{ssec:met}

In order to simulate the observed flux from our given distribution of Galactic sources, we throw anti-protons isotropically from the Earth and follow their paths until they leave the source region, while we collect the flux $\Phi(\vn)$ in each direction $\vn$ according to:
\begin{equation}
\Phi_\oplus(\vn) \propto \int_\oplus^\text{end} 
\rho(l) \text{d}l \, . \label{backt}
\end{equation}
To generate the maps of expected flux we followed about 65,000 trajectories, roughly one per degree in each direction. For each map we assumed monochromatic sources. Once the flux in each point is computed, the map is smoothed with a gaussian profile of $3^\circ$ radius to erase artificial numerical effects, and normalised. To better visualise the flux distribution we then divided each map into $5$ regions (bands) by the lines of constant {\em flux density}. The boundary values are chosen in such a way that each band contains the same fraction ($1/5$th in our case, for there are 5 bands) of the total flux.  The lighter (darker) regions correspond to smaller (larger) flux density.  The maps are in Galactic coordinates where the GC is in the centre, with Galactic longitude $l$ increasing leftwards.

At the lowest energies within our range the time spent by CRs in the galaxy becomes quite long: for events in the highest flux region at 1~EeV the maximal trajectory length can be around 100 kpc.  Fortunately, even at matter and radiation densities such as those in the Milky Way we need not worry about energy losses, for the lengths travelled by protons in a $0.3/\text{cm}^3$ density Galaxy are at least 300~Mpc due to photo-secondary production~\cite{Berezinskii} and more than 30~Mpc for p-ISM collisions~\cite{Huang:2006bp} at energies above $10^{15}$~eV. This means that in our simulations no protons will disappear along the way.

Let us stress once again that in this work we are always working in the transition regime between ballistic and diffusive propagation; true diffusion kicks in only a much lower energies, which is why it would be unsuitable to employ the diffusive approximation (and ensuing intuition).  Our signatures are due, in the first place, to the regular features of the trajectories which depart from the area of highest flux.  Now, with our method the flux is directly proportional to the line integral (along the travelled path) of the source density~(\ref{backt}).  Since then the sources are primarily distributed in the disk, CRs will pick up an appreciable flux only there; when CRs leave the disk they traverse the halo, but this part of the trajectory does not contribute to the flux, thus making the halo field essentially unimportant.

\subsection{Modification factor}
\label{ssec:spec}

Consider first the evolution of the {\em total} flux as the propagation regime changes from a ballistic to a diffusive one. To disentangle the effects of the source energy spectrum from those due to propagation, we assume in this section that the source injection spectrum is energy-independent; all the changes of the total flux with energy, therefore, result from the propagation. To avoid confusion, we call such a spectrum ``modification factor''~\cite{Berezinskii}: the physical flux is obtained by convolving it with the actual (energy-dependent) source injection spectrum.

\begin{figure}
\begin{center}
  \includegraphics[width=0.48\textwidth]{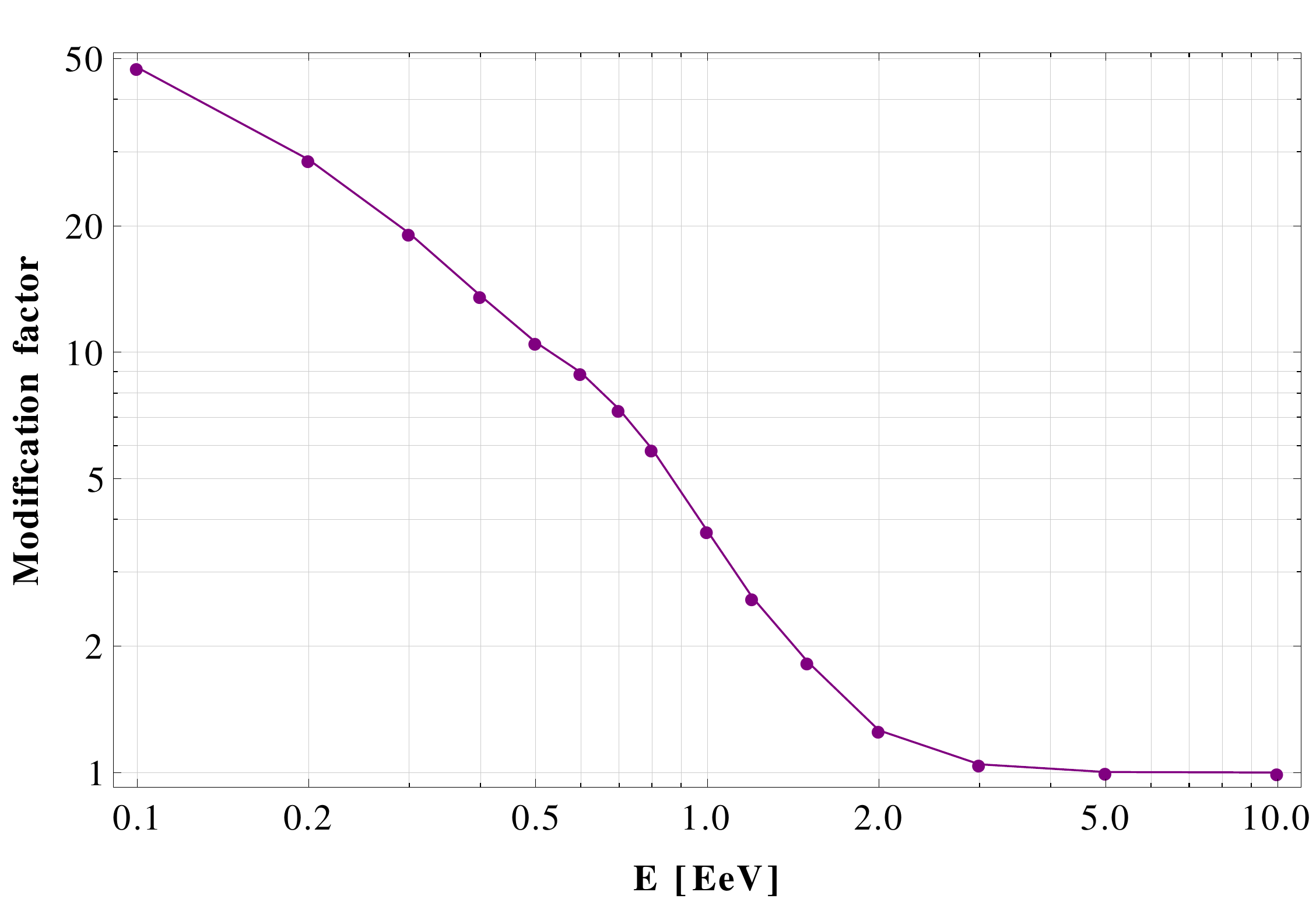}\hspace{0.02\textwidth}
  \includegraphics[width=0.48\textwidth]{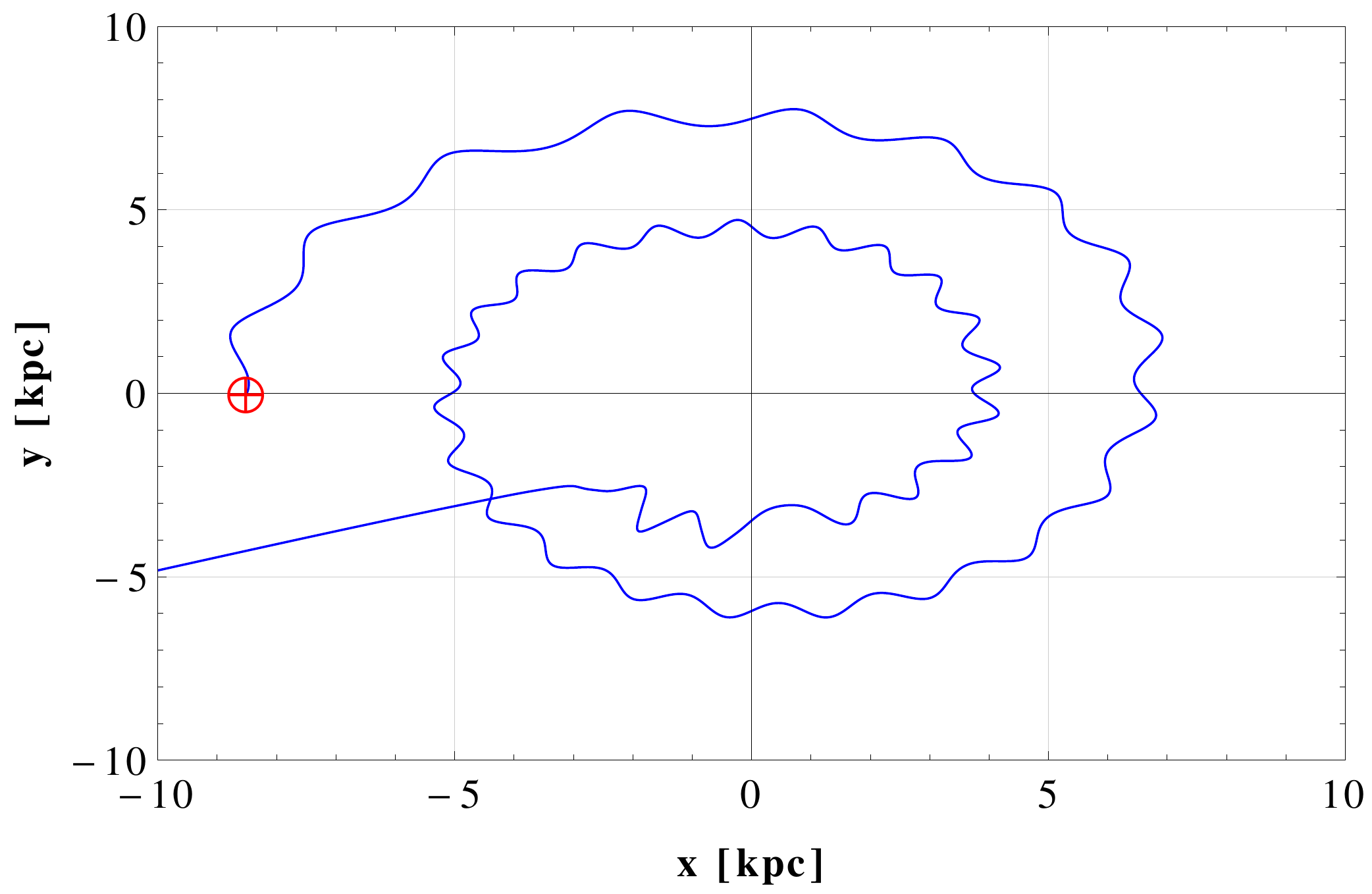}
\end{center}
\caption{Modification factor as a function of energy (left panel), and the path followed by a 1~EeV anti-proton, launched from the Earth $\oplus$ at $x=-8.5$~kpc in the direction $(l,b)=(75,-25)$, face-on view (right panel).}
\label{fig:traj}
\end{figure}

In Fig.~\ref{fig:traj} (left panel) we show the modification factor as a function of energy. At very high energies the backtracking trajectories become almost straight lines. In this limit the energy dependence of the modification factor disappears and it asymptotes to a constant; we normalise this constant to one. We see from Fig.~\ref{fig:traj} that the modification factor rapidly grows as the energy decreases. 

The mechanism behind this enhancement is evident in our backtracking method. As the energy decreases, some trajectories start to wind around the GMF lines. Since the field lines around the Earth follow the Galactic arm, these trajectories stay for a long time in the region with a high source density, so that the integral in eq.~(\ref{backt}) acquires a large value. This is illustrated in Fig.~\ref{fig:traj} (right panel) where we show one of the trajectories, at $E=1$~EeV, which gives a very high flux, since the anti-proton follows the arm spiralling inwards. 

For higher GMF strength values we have also observed an even sharper feature, due to the fact that the radius of the helix described by the particle motion becomes smaller as the particle falls inwards, since the field lines become more dense; at some point the direction of motion can even be reversed, and the CR would travel backwards along the arm again before eventually leaving.

This picture makes it clear that the flux enhancement is not a magnetic re-focussing of the flux which would otherwise arrive from a different direction, but it is an overall enhancement above the flux expected at higher energies. It also suggests that the enhancement occurs --- at least, at first -- for particular directions; because of that, a strong anisotropy is produced.

\subsection{Anisotropy}\label{ssec:maps}

\begin{figure}
\begin{center}
  \includegraphics[width=0.48\textwidth]{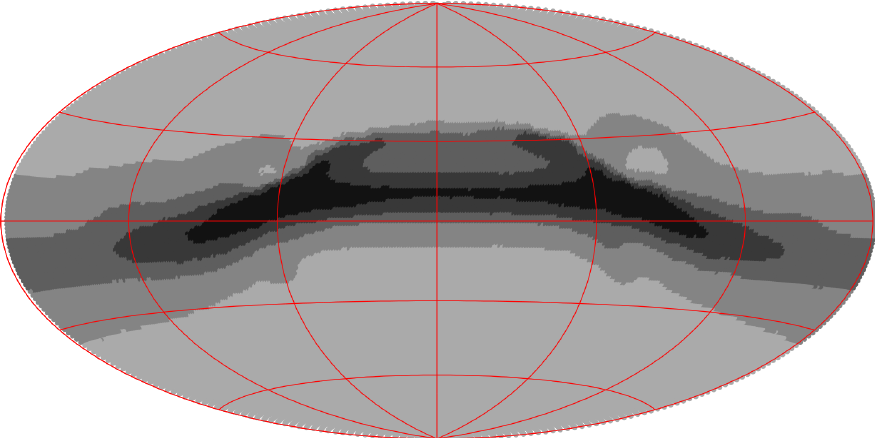}
\end{center}
\caption{Sky map of the expected CR flux from Galactic sources distributed according to~(\ref{distror}) and~(\ref{distroz}) emitting at 3~EeV.  Galactic coordinates with the GC in the centre, and Galactic longitude $l$ increasing moving leftwards.}
\label{fig:map30}
\end{figure}

Consider now the flux distribution over the sky.  In Fig.~\ref{fig:map30} we show the flux expected at the Earth from Galactic sources with injection energy of 3~EeV, at which protons propagate quasi-ballistically.  As anticipated, the map shows an image of the source region, which is a disk around the Galactic plane, distorted by the effect of the GMF.

\begin{figure}
\begin{center}
  \includegraphics[width=0.48\textwidth]{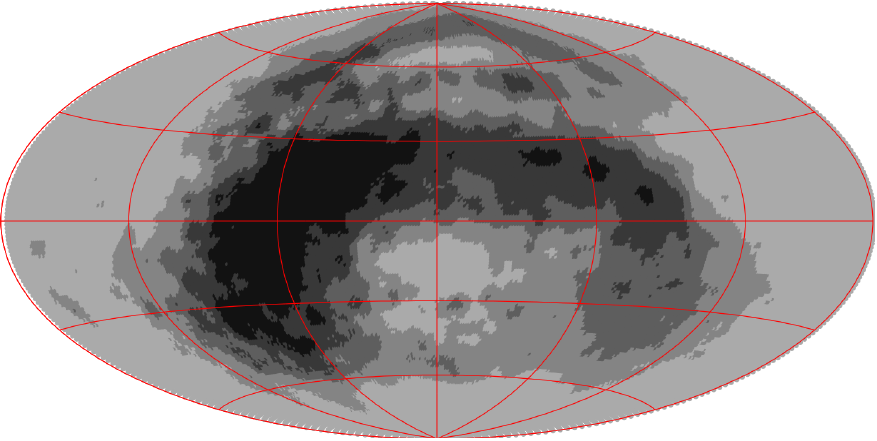}\\\vspace{10pt}
  \includegraphics[width=0.48\textwidth]{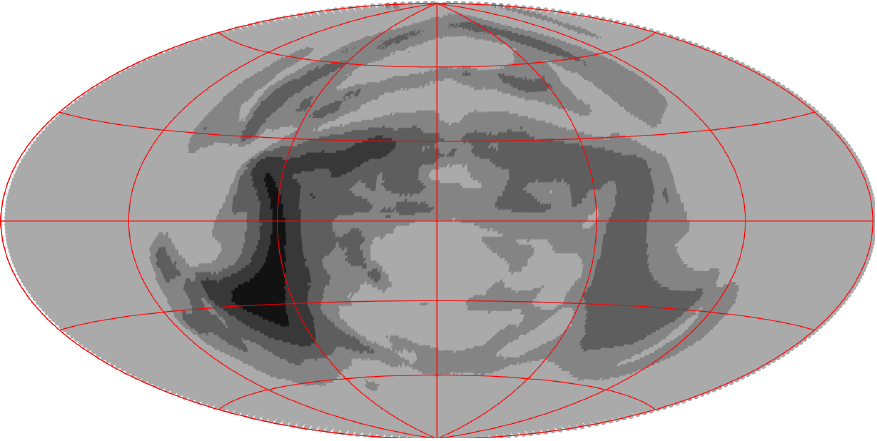}\\
  \includegraphics[width=0.59\textwidth]{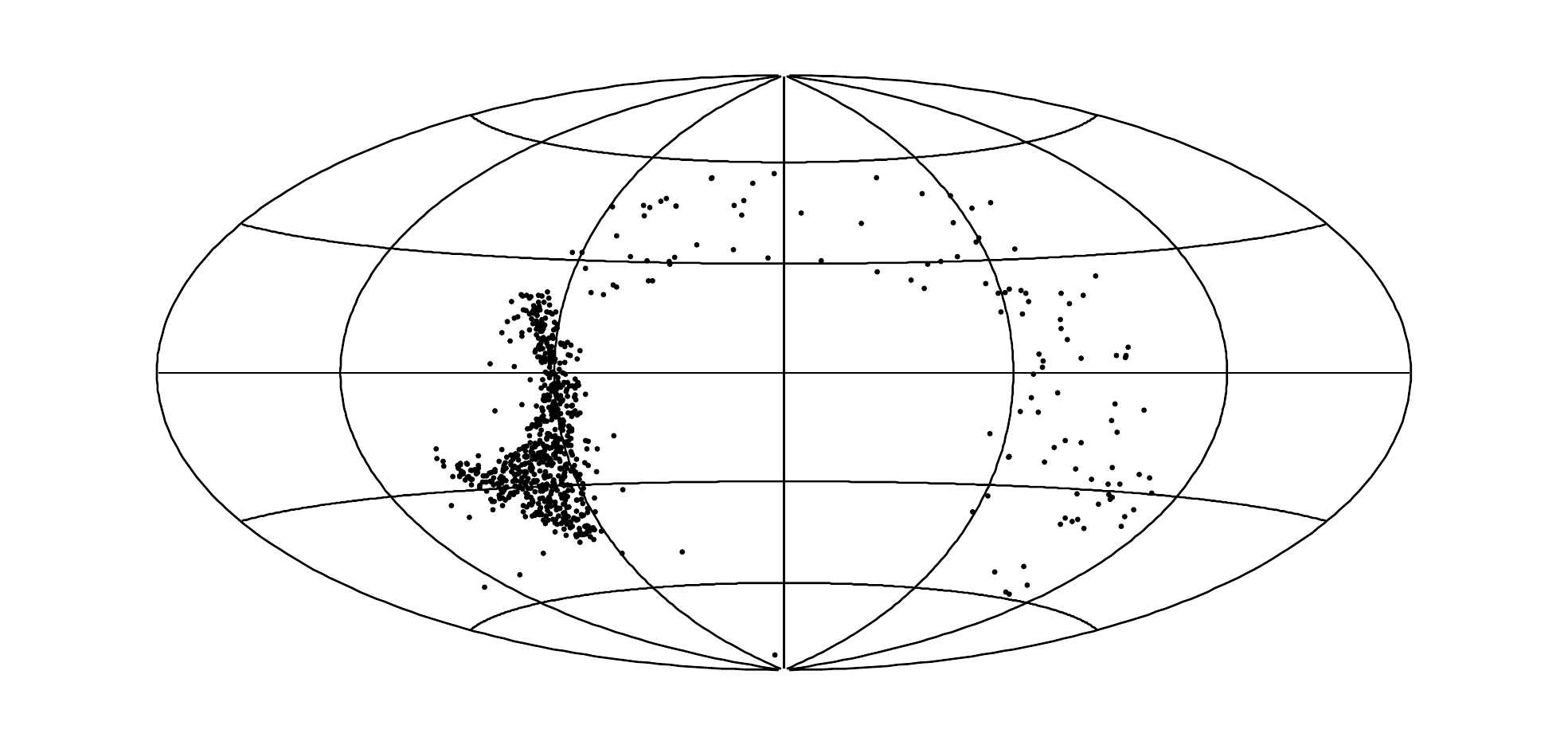}
\end{center}
\caption{Same as Fig.~\ref{fig:map30}, but for 1~EeV injection.  The top panel shows the map with the random GMF turned on, like in Fig.~\ref{fig:map30}; the middle panel is the same map but with the random component set to zero; finally, the bottom panel is the same 1~EeV skymap without turbulent GMF, obtained with the direct forward tracking method (see text).}
\label{fig:map10}
\end{figure}

Fig.~\ref{fig:map10} (top panel) is the same but now for the injection energy of 1~EeV.  The highest flux density region remains quite localised, but now its shape is not related to that of the Galactic disk.  The flux peaks around $l=70^\circ$, just below the Galactic plane, and extends roughly for about $60^\circ$ in each direction. The effects of the turbulent field are evident in the map as the jagged profiles around the bands; the deflections of the random GMF with our normalisation can amount, at this energy, to several degrees, but are still subdominant compared to regular GMF ones. Note that according to Fig.~\ref{fig:traj} (left panel), the total flux is now 4 times higher than that at 3~EeV.

It is instructive to look at what the skymap looks like when the random field is turned off, as we did in Fig.~\ref{fig:traj} for a specific trajectory.  The resulting map is shown in Fig.~\ref{fig:map10} (middle panel): the flux from a smaller region located approximately in the same direction as before (top panel) is even further enhanced compared to the average flux.  The effect of the turbulent field is thus to scatter the flux, and to randomly ``kick out'' some CRs from the arm, where they would otherwise be stuck for a much longer time, as we have seen in Fig.~\ref{fig:traj}.

Fig~\ref{fig:map10} (bottom panel) shows the result of the forward tracking method for the same 1~EeV skymap without turbulent GMF, which is meant to be a direct comparison with the skymap in the middle panel of the same figure.  Here 10,000 sources were placed in the Galaxy according to equations~(\ref{distror}) and~(\ref{distroz}), and 1000 protons were generated isotropically from each source.  The protons were followed in 1~pc steps through the GMF, and for those that approached within 50~pc of the Earth, we plotted their arrival directions in the figure.  Within the statistics, the similarity between the middle and bottom panels of Fig~\ref{fig:map10} is striking.

\begin{figure}
\begin{center}
  \includegraphics[width=0.48\textwidth]{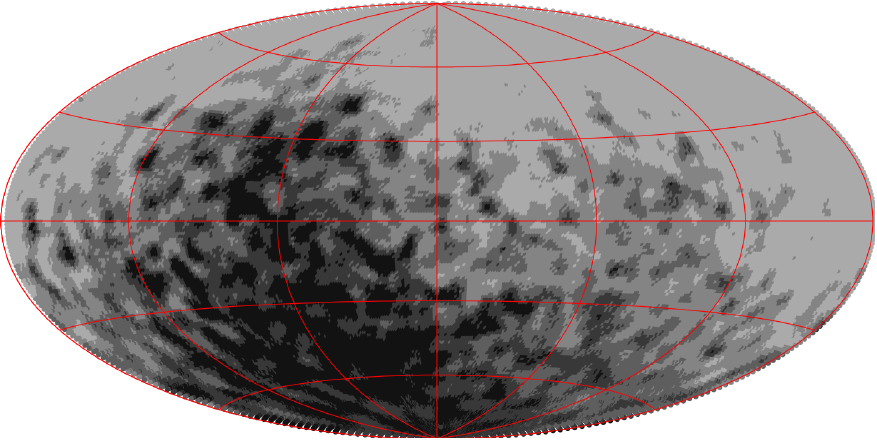}
\end{center}
\caption{Same as Fig.~\ref{fig:map30}, but for 0.3~EeV injection.}
\label{fig:map03}
\end{figure}

Finally, in Fig.~\ref{fig:map03} we show the flux at 0.3~EeV.  We can still discern the higher flux area to the left of the GC, but the map is smoothened significantly by the small-scale GMF.  The total flux continues to increase, for there are more regions in the sky for which the propagation has switched from ballistic to semi-diffusive.

As a final comment, we stress again that the qualitative picture at which we arrive is robust against the details of the source distribution, provided the latter peaks in the disk. Moreover, it should also be robust against different models of the GMF: the appearance of a special direction is a generic feature related to the local GMF (e.g., in the vicinity of the Earth) and the arm structure of the Milky Way, which is reproduced by most GMF models.

\section{Discussion}\label{sec:disc}

\subsection{A toy model}\label{ssec:fluxx}

As we have seen in the previous section, the Galactic flux of CRs has several characteristic features resulting from the diffusive-to-ballistic transition: a sharp dip feature in the spectrum, and a strong anisotropy.  In order to see how these signatures manifest themselves in a realistic situation, consider a toy model where the cosmic ray flux in the energy range around the EeV is a sum of a given --- presumably small --- fraction $f_0$ of the anisotropic Galactic flux calculated above, and an isotropic part: this component by assumption represents the extra-Galactic flux, but any heavy Galactic primaries would contribute to the flux in a very similar, that is, isotropic, way.  In the paper we use ``isotropic'' throughout, for clarity, but this observation should be borne in mind.  This approach is justified by the fact that the limits on the spatial anisotropy in this energy range are at the percent level~\cite{Aab:2014ila,ThePierreAuger:2014nja}, and because the anisotropic features of the Galactic part are quite prominent.  The fraction $f_0$ is the parameter that we want to constrain from above in the best possible way.

For the dominant, isotropic part we assume a simple power law spectrum with the spectral index $\gamma=-3.25$ as inferred from the HiResII data~\cite{Abbasi:2007sv} for energies between 0.1~EeV and 4.5~EeV --- we employed HiResII data since it nicely covers the entire energy range we are interested in\footnote{While the Telescope Array and Pierre Auger collaborations made available much more recent data, we decided to stick to HiResII data for the following reasons.  First, all three experiments' spectra agree beautifully around the ankle energy, once an energy scale increase of about 10\% is made to the Auger spectrum.  HiRes and TA spectra have essentially identical energy scales.  The lowest energies of the published HiResII, TA, and Auger spectra are log(E/eV) = 17.2, 18.2, and 17.5, and they all extend to energies higher than are relevant for this paper: the HiResII spectrum covers the largest energy range, and it is therefore logical to use that spectrum.}.

To model the Galactic part we proceed as follows.  First, we fit the modification factor of Fig.~\ref{fig:traj} with a broken power law at low energies: $0.1~{\rm EeV} < E < 1.5~{\rm EeV}$ --- this gives us a knee-like spectrum with bending energy of $E_\text{b} \simeq 0.64$~EeV and slopes $p=-0.95$ and $q=-1.86$ before and after the break\footnote{Notice that this fit is given for practical purposes only: it is not meant to describe the actual data nor an actual truly diffusive propagation; in fact, heavier elements springing from the Galaxy would flatten the slope of the total flux at lower energies, since they would keep diffusing up to a much higher energy.  Analysing such scenarios is beyond the scope of this paper so we will not pursue this any further.}.  We then adjust the low-energy Galactic slope to match the slope of the HiResII best-fit, that is, $\gamma=-3.25$; this gives us a fitted ``injection'' with power $\gamma_\text{inj}=-2.30$.  In this way, by definition, the ratio of the Galactic to extra-Galactic fluxes in the low-energy range equals $f_0/(1-f_0)$; moreover, the observed spectrum at $0.1~{\rm EeV} < E < 0.64~{\rm EeV}$ is now reproduced independently of the Galactic fraction $f_0$.

At energies higher than $\sim 2$~EeV the modification factor levels to 1, so the Galactic contribution has the overall slope of $-2.30$ and starts to win over the extra-Galactic one as the energy grows.  In order to not immediately contradict observations, we assume that the Galactic injection spectrum has a cutoff somewhere above $E\gtrsim 2$~EeV, so that the Galactic part is turned off above these energies.  This fully determines the cosmic ray flux as a function of energy in our toy model for a given value of a single free parameter, the Galactic fraction $f_0$.  Let us remark that in fact this cutoff does not appear anywhere in our tests.  The latter would be relevant only if we were to extend our spectrum beyond 10~EeV, which we never do: in the energy region we work thus the cutoff is not needed; we mention it here only for completeness.  In any case, the cutoff would not make any significant difference in our results because the main features (anisotropy and spectral drop) appear at much lower energies.

\begin{figure}
\begin{center}
  \includegraphics[width=0.8\textwidth]{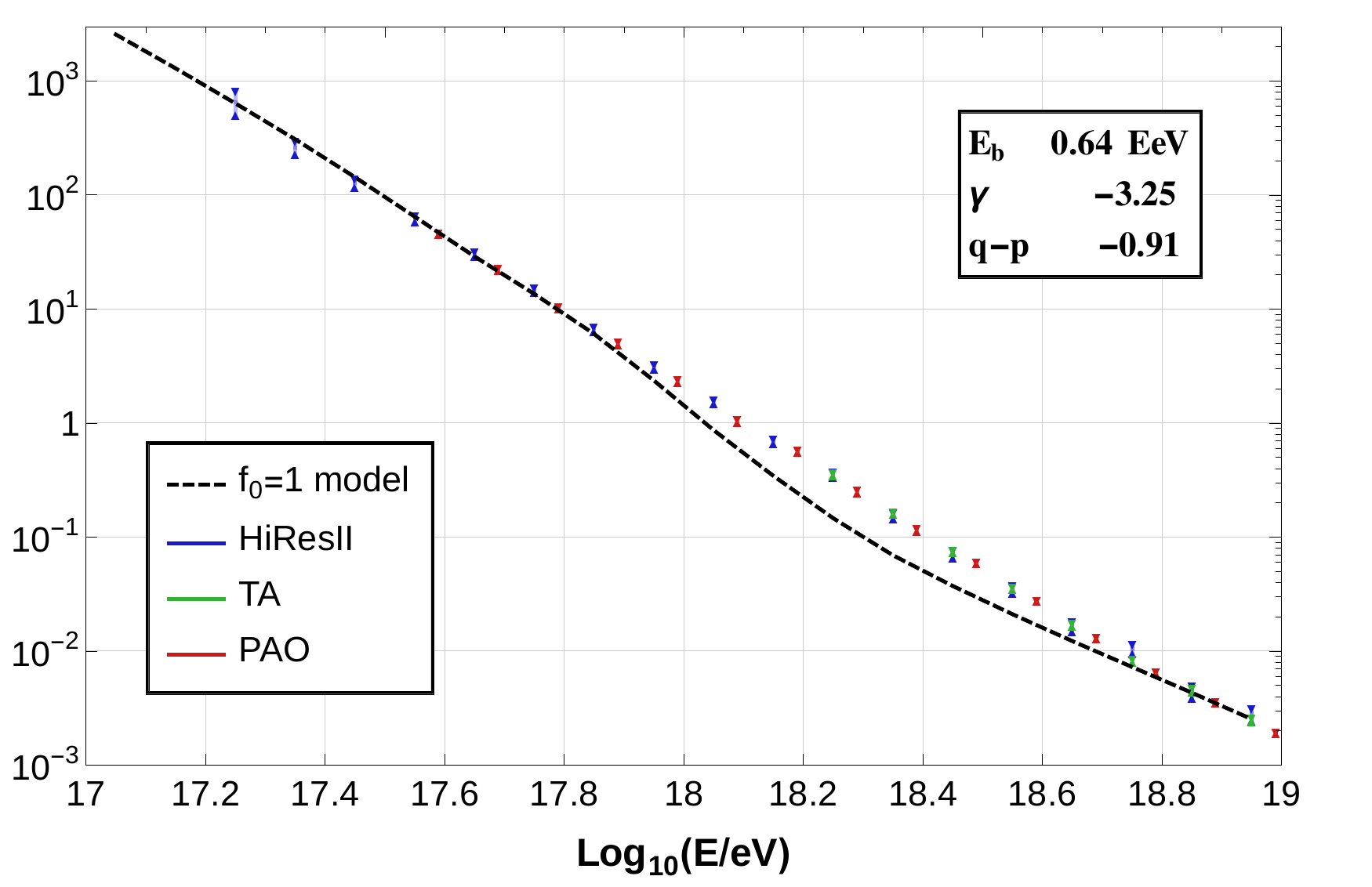}
\end{center}
\caption{HiResII (blue), Telescope Array (green), and Pierre Auger (with energy rescaled by 10\% --- red) spectra, compared with the $f_0=1$ simulated spectrum (dashed black line) from Galactic sources distributed according to~(\ref{distror}) and~(\ref{distroz}) --- no isotropic extra-Galactic flux is shown here.}
\label{fig:chi2}
\end{figure}

To illustrate the result of our procedure, in Fig.~\ref{fig:chi2} we show the HiResII spectrum (data points with errors) alongside the Telescope Array~\cite{Ivanov:2015} and Pierre Auger (with energy rescaled by 10\%)~\cite{ThePierreAuger:2013eja}, and the toy model obtained with $f_0=1$ Galactic flux (dashed).  The spectral features around the EeV mark are quite prominent.  Once again, Fig.~\ref{fig:chi2} is not meant to be an actual model fit for the data; it serves the purpose of understanding the physical effects at play.  Later we will present an order of magnitude estimate for the sensitivity of current data to such features.

Notice that, for $E > 0.64$~EeV, where the modification factor deviates from the $\gamma=-3.25$ power-law behavior, the fraction of Galactic cosmic rays becomes energy-dependent, that is, $f \equiv f(E)$.  The fraction $f(E)$ equals $f_0$ at low energies and varies in a way determined by the modification factor of Fig.~\ref{fig:traj} from $f_0$ to a negligible value.  As is clear from the discussion in Sec.~\ref{ssec:spec}, this qualitative behavior of $f(E)$ is largely model-independent modulo a possible overall shift in energy determined by the exact value of the Galactic magnetic field in the vicinity of the Earth.

Having specified the model, we can now discuss its observational signatures and the possible ways to best constrain the model parameters.  As the Galactic and extra-Galactic contributions differ by their energy dependence and sky distribution, below we compare three methods: a $\chi^2$ test for the spectral features we just discussed, a harmonic analysis in terms of a multipole expansion (both in one and two dimensions --- a similar approach can be found in~\cite{Pohl:2011ve,Giacinti:2011ww}), and a Kolmogorov-Smirnov (KS) test for the distribution of the flux along a given coordinate.

A final remark before discussing the tests individually.  In a realistic scenario the flux which an experiment would see is collected from all energies above a given cutoff.  While the maps presented above in Figs.~\ref{fig:map30}--\ref{fig:map03} correspond to monochromatic injected particles, they can be easily generalised to a full injection spectrum of particles with energies above certain threshold $E_{\rm th}$.  Because the spectrum is steeply falling, the effect of this change is minor.  Nevertheless, when discussing the statistical tests below, we generate the sky maps in this way.

Finally, an important parameter is the number of events $N_{\rm ev}$ which in a realistic situation depends on the exposure of the experiment and on the energy cut $E_{\rm thr}$.  We fix this to $N_{\rm ev}=5\times 10^4$, a realistic number for present experiments and energies around $1$~EeV.  This is the main parameter that determines the sensitivity of the tests.

\subsection{Spectral features}\label{ssec:chi2}

The first observable we can employ to tell Galactic from extra-Galactic CRs
apart is the total flux.  The procedure for obtaining the model flux for a
given $f_0$ was described above in Sec.~\ref{ssec:fluxx}. We can then build
the actual ``simulated'' spectrum, which is finally binned in energy bins of
width $\log_{10} E/\text{eV} = 0.1$.  In Fig.~\ref{fig:chi2} one can easily
see the departure from the HiResII data points around 1~EeV and beyond: this is the
feature we exploit in our test.

In order to constrain the Galactic fraction $f_0$ we generate, as mentioned previously, a set of $N_{\rm ev}=5\times 10^4$ events with energies above a given threshold distributed in energy according to the HiResII best-fit spectrum.  As an example, we consider the value $E_{\rm thr} = 0.9$~EeV for which the anisotropy is typical of the ballistic-to-diffusive transition region, and is in some sense equivalent to the monochromatic $1$~EeV map, as it can be seen by shifting the Galactic spectrum in Fig.~\ref{fig:chi2}.  These events were then binned and compared to our model spectra for different values of $f_0$ by means of a $\chi^2$ test.  For each given fraction $f_0$ we generated 1000 trial sets, and looked for the $f_0=f_{0,\,{\rm min}}$ which can be ruled out at $3\sigma$~CL in 68\% of trial sets.  This turned out to be $f_{0,\,{\rm min}} \approx 0.12$; thus, a Galactic fraction of about ten percent can be confidently ruled out based on current experimental data.  This result is similar to that obtained in the anisotropy tests, as we will see shortly.

\subsection{Anisotropy signatures}\label{ssec:harm}

As any angular distribution on the unit sphere, the CRs flux $\Phi(\bn)$ in a given direction $\bn$ can be decomposed in spherical harmonics $Y_{\ell m}(\bn)$ as 
\begin{equation}
\label{ylm}
\Phi(\bn)=\sum_{\ell\geq0}\sum_{m=-\ell}^\ell a_{\ell m}Y_{\ell m}(\bn) \, .
\end{equation} 
From the $a_{\ell m}$ coefficients one can define a direction-independent
angular power spectrum $C_\ell$: 
\begin{equation}
\label{cell}
C_\ell=\frac{1}{2\ell+1}\sum_{m=-\ell}^\ell \left|a_{\ell m}\right|^2 \, .
\end{equation}
The quantities $4\pi C_l$ characterise the amplitude (squared) of a relative flux deviation from the isotropic value.  Practically, however, as discussed at length in~\cite{Aab:2014ila}, the use of $C_\ell$ is feasible (and desirable) only when the entire sphere (the sky) is covered, for example by jostling data from several ground-based experiments.  

Alternatively, one can focus on a similar decomposition but in a single coordinate, defining the harmonic coefficients on a circle as 
\begin{equation}
\label{cndef}
c_n \equiv \frac12 \int\text{d}\bn \Phi(\bn) Y_n(\phi) \, , 
\quad \text{with} \; Y_n(\phi) \equiv \frac{1}{\sqrt{2\pi}} \, 
\left\{ \sqrt2\cos{n}\phi, \, 1, \, \sqrt2\sin|n|\phi \right\}
\end{equation}
for $\{ n>0, \; n=0, \; n<0\}$, respectively.  These one-dimensional coefficients can be measured in a single experiment without making any assumption about the flux. Similar to the 3-dimensional case, the relative deviation from isotropy (squared) is characterised by the direction-independent combination $c_n^2+c_{-n}^2$.

Now, because the spectrum is steeply falling, the higher is $E_{\rm thr}$, the smaller is the available number of events and thus the smaller the sensitivity of statistical tests.  On the other hand, the degree of anisotropy of the Galactic contribution increases.  Therefore, one generally seeks a compromise value of $E_{\rm thr}$.  Such a study is not possible without fully specifying the model and the dataset, for it implies scanning over $E_{\rm thr}$, which in turn implies knowing the exact dependence on $N_{\rm ev}$, so we do not attempt it here.  We will not be able, therefore, to compare the sensitivity of different methods in constraining $f_0$ in this section.

A simpler question, however, can be addressed within our toy setup --- that of constraining the fraction $f(E_{\rm thr})$ for a given value of $E_{\rm thr}$, which again we take to be 0.9~EeV.

The simulation is performed as follows.  Once $E_{\rm thr}$ is fixed, the sky distribution of the Galactic flux is calculated as described in Sect.~\ref{sec:res}, and the total flux is obtained by adding $f\equiv f(E_{\rm thr})$ parts of this Galactic flux and $1-f$ parts of an isotropic one.  Note that because of the presence of the random GMF component, this map contains fluctuations of the flux which vary from trial to trial.  We make only one realisation of the map and use it in all the tests.  We then study the sensitivity of the observables $C_l$ and $c_l$ to non-zero values of $f$.  We limit our analysis to $C_1$, $C_2$ and $c_1$, $c_2$ as these are expected to be dominant.

Moreover, we run direct KS test with respect to Galactic longitude and latitude, again for varying, small, $f$.  This test is designed to determine whether two sets of events are drawn from the same underlying distribution: in the case at hand we are interested in the spatial events distribution.  Now, the peculiar shape of the anisotropy around 1~EeV can be understood in an intuitive way: what we expect to see is an excess of flux while scanning the sky in Galactic latitude, which corresponds to the Galactic arm enhancement, and a deficit towards the Galactic centre while scanning along Galactic longitude.  These features would be very easy to look for and to quantify in a given experiment.  The meaning of the KS test in this respect is to simply formalise this search in a more rigorous way.

Consider first the multipole tests.  For a finite number of events, the inferred multipole components fluctuate depending on the exact event positions; the accuracy of the multipole measurement, for a given $N_{\rm ev}$, is limited by these fluctuations.  For reasonably small values of $f\lesssim 0.2$ and the adopted $N_{\rm ev}$, the distribution of multipole values, both 2- and 3-dimensional, is well described by a Gaussian with the mean that is linear in $f$, and the spread that is $f$-independent and scales with $N_{\rm ev}$ as $\propto \sqrt{N_{\rm ev}}$.  We have checked these dependencies numerically.  Thus, we only need to calculate these numbers for some particular values of $N_{\rm ev}$ and $f$. In the case of the KS test the simulations were done for varying $f$.

\begin{table}
\begin{tabular}{l|c|c|c|c|c|c}
test & $l=1$ & $l=2$ & $n=1$ & $n=2$ & KS$_{\rm lon}$ & KS$_{\rm lat}$ \\\hline\hline
$\sqrt{4\pi C_l}$ & $0.020$ & $0.018$ & --- &  ---  & --- &  --- \\
$\sigma_l$ & $0.0027$ & $0.0016$ & --- &  --- & --- &  --- \\ \hline
$\sqrt{c_n^2+c_{-n}^2}$ & --- &  --- & $0.012$ & $0.007$   & --- &  --- \\
$\sigma_n$  & --- &  --- & $0.0018$ & $0.0017$   & --- &  --- \\ \hline\hline
$f_{\rm min}$ &  $\bf 0.06$ & $\bf 0.04$ & $\bf 0.055$ & $\bf 0.09$ & $\bf 0.07$  & {$\bf 0.09$}
\end{tabular}
\caption{\label{tab:statpower} The comparison of statistical tests in their ability to rule out non-zero values of $f$ given that the data follow isotropic distribution. The columns $l=1,2$ correspond to the 3-dimensional dipole and quadrupole, respectively.  The columns $n=1,2$ correspond to the first two 2-dimensional Fourier coefficients~(\ref{cndef}).  The values of $C_l$ and $c_n$ are calculated for the case $f=0.1$.  The quantities $\sqrt{4\pi C_l}$ and $\sqrt{c_n^2+c_{-n}^2}$ characterise a maximum relative deviation of the flux from an isotropic one; $\sigma_l$ and $\sigma_n$ are their standard deviations, respectively.  KS$_{\rm lon, lat}$ stand for the KS tests in Galactic longitude and latitude, respectively.  The number of events in each simulated set was $N_{\rm ev}=5\times 10^4$.  The $f_{\rm min}$ is the value of $f$ that can still be ruled out at $3\sigma$~CL in 68\% of trial sets.}
\end{table}
To compare quantitatively the sensitivity of different statistical tests we calculate, for each test, the minimum value of $f$ which in 68\% of trial event sets can be ruled out at $3\sigma$~CL.  Table~\ref{tab:statpower} shows the results of the simulations.  As one can see from the table, the sensitivities of all the tests are quite similar and such that with $5\times 10^4$ events one can constrain the non-zero fraction $f$ at the level of 5\%--10\%.  Notice, however, that the multipole tests, as well as the KS test in latitude will somewhat lose sensitivity in a realistic situation due to the nonuniform/incomplete sky coverage a single experiment can only afford.

\section{Conclusion}\label{sec:conc}

In this work we have studied the signatures of the Galactic proton
component of UHECRs with energies between 0.1~EeV and 10~EeV; this is the range where
their propagation within the Galaxy changes from diffusive to
ballistic regimes. We have approached the problem by backtracking
individual CRs through the source region (the Milky Way), and
calculated the flux as the linear integral of the source emissivity
along the trajectory. 

Thanks to this approach two generic features of the CR flux in the
transition region become evident: (i) the flux rapidly grows as the energy decreases and the
CR propagation becomes diffusive; and (ii) this growth proceeds in a strongly
anisotropic way: all the way through the transition energy range the flux is
dominated by a bright compact spot of the angular size of a few tens of
degrees. 

The growth of the flux is conveniently represented by the modification factor
that shows the flux enhancement due to particle trapping in the MF of the
Galaxy (compared to the case of zero field). Between 0.1~EeV and 10~EeV
the modification factor has two generic features, see in Fig.~\ref{fig:traj}: a hardening around
0.64~EeV, and a flattening starting from approximately 2~EeV --- by definition, the modification factor
equals 1 at high energies when the trajectories are almost straight. These features mark the onsets of the truly
diffusive and truly ballistic regimes, respectively. We have then determined the imprint of this
modification on the observed CR spectrum in a simple toy model where a fraction $f_0$ of Galactic flux is mixed with the isotropic extra-Galactic component.  
By means of a $\chi^2$ test for simulated sets of $5\times 10^4$ events each against the model prediction we were able to exclude Galactic fractions of about 10\% at $3\sigma$~CL in 68\% of the trial sets.

We have also calculated the evolution of the flux sky maps as the energy
changes across the transition region. The flux distribution in the sky presents very peculiar
features and anisotropy patterns, which depend strongly on the injection
energy of CRs. These are easy to understand in terms of our backtracking
approach. At high energies the maps reproduce the highly anisotropic source
distribution, as they should, since the propagation is approximately ballistic. As
the energy lowers, a new anisotropy pattern arises.  Some of the backtracking
trajectories start to spiral around the lines of the local regular GMF: they thus
stay a long time in the disk where the density of sources is high. The flux
from the directions corresponding to these trajectories is strongly
enhanced: as a result a special direction, with higher flux, appears in the sky.
Finally, at even lower energies most of the trajectories become
trapped and the anisotropy is washed away (but the total flux keeps increasing): this corresponds to the onset of
the truly diffusive propagation.

The anisotropy described above is a clear signature of the Galactic
sources; it can be searched for in the current CR data by means of 2-
and 3-dimensional harmonic analyses, as well as through direct KS
tests along a particular direction.  Our simulations show that all
these methods yield similar results, that is, we are 
sensitive to Galactic flux fractions around the 5\% to 10\% level for an
energy cutoff of 0.9~EeV. This means that, based on the anisotropy it
produces, a 5\% to 10\% Galactic flux at this energy can be quite confidently
rejected based on current data.

Both qualitative features --- the flux change and the anisotropy --- are robust with respect to model details such as exact distribution of sources in the Galaxy, provided the conditions described in Sect.~\ref{sec:met} are satisfied, and details of the Galactic magnetic field. The latter is quite evident since the anisotropic flux enhancement is traced back to the trajectories which spiral around the GMF lines along our galactic arm. The same picture would arise in any model where the GMF roughly follows the arm within several kpc from the Earth. The strength of the magnetic field determines the energy at which the transition occurs.

The statistical tests we have presented in this work are ready to, and should, be applied to the actual data, in order to finally find the energy at which the isotropic extra-Galactic CR flux wins over the anisotropic Galactic one.

\subsection*{Acknowledgements}

FU was supported by the grant IUT23-6 and by the EU through the ERDF CoE program. He would like to thank the Service de physique th\`eorique at the Universit\'e Libre de Bruxelles for hospitality while this project was conceived, developed, and completed.  FU was, and PT is supported by IISN project No.\ 4.4502.13 and Belgian Science Policy under IAP VII/37.  DI and GT are supported by the U.S.\ NSF grants PHY-1404502 and PHY-1404495.

\section*{Appendix: Simulating the Galactic magnetic field}

\subsection*{Regular GMF}

In all our simulations we employed the large-scale field of~\cite{Pshirkov:2011um}.  This model consists of two components of the GMF, a disk field and a halo field.  In heliocentric coordinates with the $x$-axes looking away from the GC, and the $z$-axes pointing towards the North, the disk field has radial and azimuthal components $B^D_{\theta} = B^D(r,\theta,z) \cos{p}$ and $B^D_{r} = B^D(r,\theta,z) \sin{p}$
\be
B^D(r,\theta, z)=B^D(r)\left|\cos{\left(\theta-b\ln{\frac{r}{R_{\odot}}} +\phi\right)}\right|\exp{(-|z|/z_0)} \, , \label{B_axisym}
\ee
or
\be
B^D(r,\theta,z)=B^D(r)\cos{\left(\theta- b \ln{\frac{r}{R_{\odot}}} +\phi\right)}\exp{(-|z|/z_0)} \, , \label{B_bisym}
\ee
where $z_0 = 1$~kpc, $p$ is the pitch angle, for the models dubbed ``axisymmetric'' ($p=-5^\circ$) and ``bisymmetric'' ($p=-6^\circ$), respectively.  Here $\phi=b \ln\left(1+d/R_{\odot}\right)-\pi/2$ where $d=-0.6$~kpc is the distance to the first field reversal and $b\equiv 1/\tan p$.  $R_\odot=8.5$~kpc is as usual the distance to the GC.  The amplitude of the GMF along $r$ behaves as
\be
B^D(r)= \left\{
\begin{array}{ll}
\displaystyle B^D_0\frac{R_{\odot}}{r\cos{\phi}}, \quad &r>R_c \, ,\\
\displaystyle B^D_0\frac{R_{\odot}}{R_c\cos{\phi}}, & r<R_c \, ,
\end{array}
\right. \label{B_radial}
\ee
where $R_c = 5$~kpc is the radius of the central region where the disk field is assumed to have constant magnitude, and the best-fit overall magnitude is $B^D_0 = 2\;\mu$G.

The halo field instead is purely toroidal:
\be
B_{\theta}^H(r,z) = B_{0,n,s}^H \left[1+\left( \frac{|z|-z^H}{z_0^H}\right)^2\right]^{-1}\frac{r}{R_0^H} \exp{\left(1-{\frac{r}{R_0^H}}\right)} \, , \label{B_halo}
\ee
where the direction of the field is reversed in the Southern hemisphere.  The North and South field strength best fit values are $B_{0,n}^H = 4\;\mu$G and $B_{0,s}^H = - 2\;\mu$G for the axisymmetric case, and $B_{0,n}^H = - B_{0,s}^H = 4\;\mu$G for the bisymmetric case.  $R^H_0 = 8$~kpc is the radial scale of the halo, and $z^H = 1.3$~kpc defines the vertical position of the halo; as usual $z_0^H$ is the vertical scale, which could differ in directions to ($|z|<z^H$) and away ($|z|>z^H$) from the Galactic plane, which we call $z^H_{0,\text{in}} = 0.25$~kpc and $z^H_{0,\text{out}} = 0.40$~kpc, respectively.

\subsection*{Turbulent GMF}

We drew inspiration from~\cite{Haverkorn:2014jka} for our small-scale component, based on the regular model of~\cite{Pshirkov:2011um}.  The field is thus made of two separate components: for the disk field the reference strength was set at our location, that is, $r=R_\odot$ with $\theta=z=0$ --- we call this value $\hat B^D_0$; the halo random field was normalised instead at the vertical peak of the halo regular field, that is, $z^H$, with $r=R_0^H$ and $\theta=0$ --- this valus is called $\hat B^H_0$.  Given the normalisations, the field values elsewhere are simply obtained by rescaling according to the regular GMF, that is, $\hat B^D(r,\theta,z) = \hat B^D_0 B^D(r,\theta,z) / B^D(R_\odot,0,0)$ and $\hat B^H(r,\theta,z) = \hat B^H_0 B^H(r,\theta,z) / B^H(R_0^H,0,z^H)$, for the disk and halo fields, respectively.  Notice that this field is not divergenceless; however, the effect of the constraint on the divergence has the physical effect of slightly extending the coherence length of the field, and given the other uncertainties we are dealing with here, this is not expected to be significant.

The spectrum of each of the two random components is assumed to be of Kolmogorov type, or a somewhat shallower one, up to its outer edge $l_c$.  Therefore, in determining the random deflections of CRs due to this field we can limit ourselves to the two outer modes alone (i.e., the disk and halo outer modes), as smaller scales would produce a negligible effect.  In order to implement the random field we employed the following trick.  At each integration step (in our simulations we used $L=10$~pc steps), we set the final position and velocity anew by a random shift which correspond to what a field (or many fields) of given strength and coherence length would produce within our step.  We also assume that the deflections are smaller than the step size, that is, the Larmor radius at a given energy $E$ and charge $q$ in the turbulent field ($r_L = E/q \hat B$) should be much larger than $L$ for consistency.  Within these assumptions, the position and velocity shifts scale like $q (l_c^3 L)^{1/2} \hat B/E$ and $q (l_c L)^{1/2} \hat B/E$, respectively.


As we propagate proton primaries, in this 0.1~EeV to 10~EeV energy band the regular GMF plays the main part in determining the path followed by these CRs.  This can be understood directly by computing the random GMF with a preferred prescription, such as the one we have just described (see also, e.g.,~\cite{Beck:2014pma}); also, this can be seen more directly if we extract the maximal allowed random deflection for CRs directly from the behaviour of the linear polarisation of distant, extra-Galactic sources, i.e., Faraday rotation measures, see~\cite{Pshirkov:2013wka}.  The point being that the ratio of random VS regular deflections depends only on the ratio of field strengths and that of coherence lengths to some power $n\geq1/2$: assuming a random field 5 times stronger than the regular one (the most extreme situation in our model), the ratio of coherence lengths (around a pc for the random component, and at least several hundred pc for the regular one) suppresses the effects of the former in favour of those of the latter.  In this sense we are studying a different regime than in Ref.~\cite{Giacinti:2015hva}, as we are looking at the edge between diffusive and ballistic propagation.


\bibliographystyle{mnras}
\bibliography{Pulsars}

\end{document}